\documentclass[]{aa} % for a referee version
\usepackage{graphicx}
\usepackage{graphicx,color,ulem}                                % Use pdf, png, jpg, or eps§ with pdflatex; use eps in DVI mode

\usepackage[varg]{txfonts}
\usepackage{natbib} 
\bibpunct{(}{)}{;}{a}{}{,} % to follow the A&A style

\begin{document}

\title{A new dynamically self-consistent  version \\of the  Besan\c{c}on Galaxy Model}

\titlerunning{The  Besan\c{c}on Galaxy Model}

\author{O. Bienaym\'e\inst{1}  \and J. Leca\inst{1} \and A. C. Robin\inst{2}}

\institute{Observatoire astronomique de Strasbourg, Universit\'e de Strasbourg, CNRS,  11 rue de l'Universit\'e, F-67000 Strasbourg, France 
        \and
Institut Utinam, CNRS UMR 6213, Universit\'e de Franche-Comt\'e, OSU THETA Franche-Comt\'e-Bourgogne, Observatoire de Besan\c{c}on, BP 1615, 25010 Besan\c{c}on Cedex, France
 }

\date{Received / Accepted}

\abstract{Dynamically self-consistent galactic models are necessary for analysing and interpreting star counts, stellar density distributions, and stellar kinematics in order to understand the formation and the evolution of our Galaxy.}
{We modify and improve the dynamical self-consistency of the  Besan\c{c}on Galaxy model in the case of a stationary and axisymmetric gravitational potential.}
{Each stellar orbit is modelled by determining a St\"ackel approximate integral of motion. Generalised Shu distribution functions (DFs) with three integrals of motion are used to model the stellar distribution functions.}
{This new version of the Besan\c{c}on model is compared with the previous axisymmetric BGM2014 version and we find that  the two versions have  similar densities for each stellar component. The dynamically self-consistency is  improved and can be tested by recovering the forces and the potential through the Jeans equations applied to each stellar distribution function. Forces are recovered with an accuracy better than one per cent over most of the volume of the Galaxy.
}{}{


\maketitle

\section{Introduction}
The Besan\c{c}on Galaxy Model \citep{rob86a,rob86b,bie87, mar06, rey09,rob12, rob14, amo17, lag17, lag18}  has been created to model the observed Galactic star counts, to allow predicted star counts, and to give insight on  the structure, formation, and evolution of our Galaxy. It is a synthesis model that includes  essential elements of our current knowledge  of the Galactic physics. A model of this kind is a natural extension and modern generalisation of  methods  based on the equation of stellar statistics \citep{von98}. 
 Many similar models have been developed; we can cite  a few  of the most recent developments \citep{gir05,bin12,san14,bin14,sha14,vas18,sys18}. The Besan\c{c}on  model is based on a set of Galactic components for the ISM, stars, and dark matter. The  stellar density distribution is described with  stellar components, each component having different  characteristics of stars with  ranges of ages, abundances, and  radial gradients. 
The model  reproduces observed star counts. It can anticipate or predict the results of star counts  with many of the existing  photometric wide bands. 
The transformation of stellar parameters (effective temperature, gravity, metallicity) to observables (magnitude and colours) is done with semi-empirical atmosphere model grids \citep{lej97,wes02}, the so-called BaSeL libraries. For the very cool dwarfs with temperatures lower than about 4000 K, we use  the NextGen models (\citealt{all97,bar98}; see also  \citealt{sch06}).
Among some of the recent improvements, we  mention the introduction of non-axisymmetric components like the inner bar structure  \citep{rob12}. The model also includes a  detailed modelling of the extinction allowing an accurate description of observations towards directions close to  the Galactic plane \citep{mar06}.
The model is periodically updated with  the regular advances of our knowledge of  Galactic properties, stellar properties, and luminosities \citep{sch06,cze14,mor17,amo17,lag17,lag18}, binarities, abundances, stellar kinematics, and dynamics.
\\
Here we are concerned  with the improvement of the dynamics of the model, and  we mention the preceding introduction of  the kinematical properties of stars \citep{rob87},  and the introduction of a dynamical consistency in the solar neighbourhood \citep{bie87}. This dynamical consistency relates the thickness of the stellar components to their vertical velocities through the vertical gravitational potential close to the solar Galactic radius $R_0$.
In \cite{bie87}  the dynamical consistency is restricted to the solar neigbourhood. 
A similar  Galactic model, but with an inner bar and  a dynamical consistency  restricted to the solar neigbourhood, is being produced (Fern\'andez-Trincado et al., in preparation).

 It has been known for decades that  globally dynamically consistent models of the Galaxy can be built \citep{mcg90,ken91,fam03}, but only recently have practical tools  been developed using action integrals \citep{bin12} or energy-integrals  that   achieve  excellent accuracy \citep{bie15}.
In the present paper the consistency is extended to a much broader space, nearly 95 \% of the volume occupied by the stellar discs, with the exception of the very inner part of the Galaxy model.
 Our dynamical model assumes axisymmetry, a hypothesis  not satisfied in the inner part of the Galaxy.
 
To model the distribution of the stellar disc components, we define  general distribution functions depending on three integrals of motion, the energy, the angular momentum, and a third approximate integral. They are derived from the Shu distribution function (DF).
\\

The arrival and publication of Gaia data offer the possibility to probe a wide volume of the Galaxy with  unprecedented accuracy. It implies that the quality of the modelling must be realised  with regard to these exceptional data. 
Owing to the exquisite accuracy of  observations, many Gaia data can be directly inverted to recover 3D positions and velocities of stars with negligible bias. 
This  makes  it simpler to perform the analysis within an extended neighbourhood of the Sun, and it avoids the difficulties of applying classical analysis based on the apparent magnitudes, proper motions, radial velocities, or spectrophotometric distances.
In particular,  Gaia data will allow us to measure with a never achieved precision the Galactic potential over an extended domain. Determining accurately the mass contribution of the visible matter, a challenge even with Gaia data, and subtracting it from the total dynamical mass computed from the potential,   will allow us to draw very precisely the dark matter density distribution.
One way to achieve this goal consists in building  dynamically self-consistent Galactic models that relate the  kinematics and the number density for each stellar population through the collisionless Boltzmann equation under the hypothesis of stationarity.
Such models are based on explicit distribution functions  \citep{tin13,bin14,bov15,bie15,bin16,vas18}  and these available models  follow  identical  approaches   \citep[see a compilation by][]{san16}  using  St\"ackel fits and  fitting orbits with  actions at the exception of \cite{bie15} who used analytic integrals of motion. 
With the exception of \citet{sanbin15} who develop a triaxial model, 
these models do not  yet include  non-axisymmetric effects,   for instance related to a triaxial dark matter halo or to elliptical discs.
Naturally, other approaches are possible, similar to  these based on N body simulations \citep{che15}  and    also attempts to numerically fit more general integrals of motion from torus reconstruction \citep{pre82,mcg90,rob93,kaa94a,kaa94b,bie13}.
Finally, we note that building a stationary model is  a first step in order to determine the amplitude of non-stationarity  and   can be used  as  a reference  state to analyse  non-stationarities like perturbations.
\\

In this paper, we detail the way we generalise the self-consistent dynamical modelling for an axisymmetric version of the  Besan\c{c}on Galaxy Model (BGM}). The  accuracy of this method is quantified in \citet{bie15}. It is shown that  the  2014 version of the BGM \citep{rob14} is not far from the dynamical self-consistency  and that  the majority of orbits  are  conveniently  described in the frame of St\"ackel potentials. An important test consists in recovering the full Galactic potential from the Jeans equation using   distribution functions  built with the mean of integrals of motion. Applying our method, the vertical and radial forces and the total mass density are recovered with an uncertainty smaller than one percent  in a large volume of the Galaxy.\\

Our approach of the dynamical self-consistency is formally identical to that  developed by \citet{san16}     also using the formalism of St\"ackel potential. A significant difference is that we  use an explicit analytic expression of the integrals, but not the numerically integrated actions. Another aspect is that our analytic expressions are  rapidly evaluated.
\\

This paper presents the different steps used to build a dynamically consistent version of the Besan\c{c}on model. The first step is the determination of the gravitational potential based on usual methods, and then the determination of the approximate third integrals for any point in the phase space. It follows with  the description of distribution functions for disc stellar populations based on a generalisation of the Shu DF.  Then we present a comparison of the previous BGM2014 version with this new one.

\section{Potentials}

The core of the dynamical modelling is the calculation of the  gravitational potential and forces. Their  determinations do not present critical difficulties in the present context of components without a central cusp and with a null density  at large distances. A  review of some numerical methods is given in Vassiliev (2018). In the different versions of the BGM \citep{bie87,rob12,rob14},  the components either have  an explicitly known potential-density pair or, in the case of  the  ellipsoidal density distributions with Einasto profiles,   the potential is determined with a single  integration. Comparing our new version, presented here, with the BGM2014 version \citep{rob14},  three of the analytical density components of the BGM--the youngest
stellar disc, the ISM, and the stellar halo are not modified--but  we change the analytic expression for the dark matter to allow its flattening. In the previous BGM versions the  stellar densities are modelled with Einasto profiles or with modified double exponential profiles. Within this new version, stellar discs, which are dynamically consistent,  are numerically determined and tabulated on a $(R,z)$ coordinate grid. Furthermore, the density of stellar components are set null beyond a cut-off radius, $R_{cut} \simeq$  15\,kpc \cite[see][]{amo17}. We also introduce a vertical cut with a null density  beyond $z_{cut}=\pm$4 kpc (the dynamical modelling of thin discs gives a density at 4 kpc four orders of magnitude or more smaller than at $z=0$ and  thus is numerically negligible).
We leave  the youngest analytical  disc unchanged since this component is not in a stationary equilibrium state and does not need a stationary dynamical  modelling.
\\

Potential and forces must be determined accurately and  the numerical computation must be fast enough. We  consider that  this implies   relative errors on forces   smaller  than three thousandths   everywhere.  Within the disc, this is sufficient  to accurately distinguish  the amount of dark matter  from  the visible matter. To compute the gravitational potential we  consider the total density from  the Galactic components having a cut-off radius (stellar discs and ISM). Other components,   e.g. dark matter and the  stellar halo,  are treated separately.

We solve the 3D Poisson equation in cylindrical coordinates in the axisymmetric case
\begin{equation}
\Delta \, \phi(R,z)= 4 \,\pi \, G \, \rho\,(R,z) 
\end{equation}
using a finite difference algorithm.

The density is discretised on a grid with $2^{N}$ points along the $R$  and $z$ directions, respectively, from 0 to $R_{max}=61$ kpc and from 0 to $z_{max}=10$ kpc. The Poisson equation is solved with  boundary conditions (BC) defined along two lines $\phi(R_{max},z)$ and $\phi(R,z_{max})$. They are numerically determined with the equation

\begin{equation}
\phi({\bold{x}}) = -G \iiint   \rho({\bold{x'}}) \times \frac{1}{|{\bold{x}}-{\bold{x'}}|}\,  {\rm d}^3 x'
\end{equation}

Significantly higher values of $R_{max}$ and  $z_{max}$ are chosen   than the outer cut-off radius of the density distributions to avoid the singularities of the Green's function for gravitational potential and to minimise  its variation between neighbouring points of the grid.  The density is evaluated and the integration performed applying the Simpson rule on a grid with steps $\Delta x=\Delta y$= 50 pc and $\Delta z$= 40 pc.

The numerical integration of Eq.\,1 is performed  iteratively and  to reduce the computing time $N$ is   progressively increased from 5 to 7 (the number of grid points passing from 32$\times$32 to 128$\times$128 with final steps  d$R$=480 pc and d$z$=80 pc). The potential and forces outside of the grid points are interpolated with   a  bicubic spline  that ensures the continuity of the derivatives at  positions on  grid points. The final accuracy depends on the different grid sizes, the grid to sample the density distribution, the one used to compute the BC, and the one to solve the Poisson equation.  The  potential is  used to determine the stellar distribution functions (see Section 4) and the forces  to compute the orbits. Forces are  also needed to test the self-consistency of the dynamical model  by checking the numerical exactness of the Jeans equations. This last test is mandatory if we want to accurately determine the total Galactic mass distribution through the density and kinematics of stellar populations \citep{bie15}.  Applying the dynamical consistency to the BGM2014 version  \citep{bie15}, we  obtain an  accuracy better than three thousandths  over a wide volume. A better accuracy of one thousandth can be obtained using a 256x256 grid to solve the Poisson equation, but with a computing time that is a  factor of 30 times  longer.
\\

In the previous versions of the BGM the  spherical dark matter halo potential is given in spherical coordinates by

\begin{equation}
 \phi(r)= -4 \pi G \rho_c  r_c^2 \left[ \,1-\frac{1}{2} \log(1+r_s^2)-\arctan (r_s)/r_s  \, \right] 
\end{equation}with $r_s=r/r_c$.
This corresponds to a density distribution characterised by its core radius $r_c$ and central density $\rho_c$
 
 \begin{equation}
 \rho(r)= \rho_c \,\frac{r_c^2}{r^2+r_c^2}
 \end{equation}providing a flat rotation curve at large radius.

We modify this dark halo potential by replacing $r^2$ with $R^2+z^2/q^2$ in cylindrical coordinates to allow a halo flattening. The resulting density remains positive if  $q>0.5$ and thus is also positive  for usually accepted values of the potential flattening.

Adding  visible and  dark matter components,  the dark matter core radius $r_c$ and  central density $\rho_c$ are adjusted   in order to reproduce the observed rotation curve  \citep{sof12}.  The flattening $q$  of dark matter halo   is adjusted to reproduce the known local mass density and local dark matter density \citep{bie14}.
\\

\section{St\"ackel fit to a stellar orbit}

Stationary distribution functions (DFs) can be written depending on the  integrals of motion  to model the density and kinematical distribution of stellar discs.  In the case of St\"ackel potentials three such integrals of the motion are known and can  be  expressed analytically  \citep{lyn62,dez85a,dez85b}. These integrals can be used to build DFs and appropriate ones are the extension of the Shu DFs  with nearly isothermal distribution of velocities. 

St\"ackel potentials  cover a relatively large variety of potentials with  many orbits,  similar to those in realistic galactic potentials. The Galactic potential can be globally approximated with a St\"ackel potential  \citep{deb00,fam03}. However, concerning the distribution functions of stars, it is more interesting to consider separately the orbits and to find, independently for each orbit, the associated St\"ackel potential that provides the best approximate third integral.  For an axisymmetric potential, two integrals are the energy and the angular momentum, and it is always possible to express explicitly the third integral as a function of the potential (see below). This analytic expression can be used  as an approximate integral for a non-St\"ackel potential. The only free parameter $z_0$  is adjusted by  minimising  the variance of the approximated third integral along each orbit ($z_0$ defines the confocal ellipsoidal coordinate system associated to a St\"ackel potential).  In practice, we adjust only one $z_0$ value for each  family of orbits having the same energy $E$ and the same angular momentum $L_z$. Then we tabulate the  corresponding fitted function $z_0(E,L_z)$.

With the definition  given below,  the third integral is exact and null for the orbits confined in the plane. The maximum values of this third integral  are reached for  orbits close to the shell-like orbits (high $z$-vertical  extensions and   small radial extensions when they cross the Galactic plane). If $z_0$ is modified, the variance of the approximate  third integral along orbits simultaneously increases or decreases   for all the orbits with the same $E$ and $L_z$.  Since the variance  is minimum  for the   shell-like orbits \citep[see figure 3 in][]{bie15},     to determine the best fitting $z_0$, it is safer in computing time to consider only these shell orbits. Furthermore, it is sufficient to follow these shell orbits  for a short time  with a single vertical excursion out of the Galactic plane.

 Other methods to approximate a third integral were proposed and are based on the determination of the actions \citep[for a recent review, see][]{san16}. The method that is presented here is similar to what was published in \citet{ken91}, where our third integral expression (see Eq. 6 below) can be recovered by substituting their equation 14 in equation 13 and the energy by the Hamiltonian. By comparing three different methods to approximate the third integral, \cite{ken91} noted that this method is the most accurate.
 
Our method  is based on the fact that  for  an orbit passing through the point $(x,y)$ = $(\lambda_0, \nu_0)$ we associate  the potential $\phi$ to a St\"ackel potential $\phi_{Staeckel}$, and both potentials are assumed to be identical along  the two lines $\lambda=\lambda_0$ and $\nu=\nu_0$ \citep[a complete discussion of St\"ackel potentials and ellipsoidal coordinates is given in][]{dez85a,dez85b}.  This is quite similar to what is done in \citet{san14} or \cite{vas18}    before they determine numerically the actions  with a quadrature. Other ways to proceed to a St\"ackel approximation are given in \citet{san16} or \citet{vas18}. Here, we were  also able to  determine the action since with the $E$ and $I_3$  expressions, it is straightforward to deduce the equation of the section of the phase space $(R,v_R)$ and $(z=0)$.  Then the   action is obtained through a  numerical quadrature.

\paragraph{Third integral:}

We reproduce in Eq. 5 the  expression of a third integral depending on the coordinates and the potentials \citep[see][]{bie15}. Elliptical coordinate systems and St\"ackel potentials are presented in \citet{dez85a,dez85b}. One of the usual forms given for the third integral is 

\begin{equation}
\label{eq1}
I_s= \Psi(R,z)
-\frac{1}{2} \frac{L^2-L_z^2}{z_0^2} -\frac{1}{2}v_z^2
,\end{equation}
where $L$ and $L_z$ are the total and vertical angular momenta, $z_0$ a fixed parameter, $v_z$ the vertical velocity, and the function $\psi$ can be written with its potential dependence \citep{bie15}
\begin{equation}
\label{eq2}
\Psi(R,z)=-
\left[ \phi (R,z) 
-   \phi(\sqrt{\lambda},0 )\right] \, \frac{(\lambda +z_0^2)}
{z_0^2}\, ,
\end{equation}
with $\phi$ the potential and $\lambda$ one of the ellipsoidal coordinates:
\begin{eqnarray}
\label{eq3}
 \lambda  = & \frac{1}{2}(R^2+z^2-z_0^2)
+\frac{1}{2}\sqrt{\,(R^2+z^2-z_0^2)^2+4\,R^2\, z_0^2} \, .
\end{eqnarray}
We use a normalised third integral $I_3$ that varies with St\"ackel potentials from 0 for the orbits confined to the plane to 1 for the shell orbits
\begin{equation}
\label{eq4}
I_3=-\frac{I_s}{(E-E_c)} \left(1+\frac{R_{c}^2}{z_0^2} \right)^{-1} \, , 
\end{equation}
with $E$ the total energy;  $E_c(Lz)$  and $R_c(Lz)$  are respectively the energy and the radius of the circular orbit that have the angular momentum $L_z$.

The benefit of this simple formulation is that the third integral explicitly depends on the potential. Generalisation to non-axisymmetry with three planes of symmetry is  possible with a second and a third integral also written with their potential  dependence (in preparation).

\section{Distribution functions for stellar discs}

Within the BGM, the DFs represent the number density of stars within the phase space. Their characteristics, age, mass, and abundances are extensions of  the DFs with more  variables and they are defined elsewhere in the BGM model. Concerning the kinematics, a stationary distribution function of positions and velocities is a solution of the collisionless  Bolzmann equation and can be expressed as a function of the  isolating integrals of motion.  Here we extend to 3D the Shu DFs that were built to define  2D stationary exponential  discs \citep{shu69}. The 3D extension is achieved owing  to St\"ackel potentials that admit three integrals of motion. The Shu DF depends on $E$ and $L_z$. The 3D generalisation includes the vertical motions and the distribution outside of the $z=0$ disc  by adding a vertical nearly  isothermal distribution  using a third integral. This generalisation is  defined in \citet{bie99}  to analyse  local samples of Hipparcos stars and is also used to determine the local density of dark matter, modelling stellar samples of RAVE stars \citep{bie14}.\\

In the case of St\"ackel potentials a generalised Shu DF can be written as
 \begin{equation}
f(E,L_z,I_3)=g(L_z) \exp \left (-E_R /\tilde{\sigma}_R^2 -E_z /\tilde{\sigma}_z^2 \right)
 \end{equation}
with 
 \begin{equation}
E_R= (E-E_c) \, (1-I_3)
 \end{equation}
and 
 \begin{equation}
E_z= (E-E_c) \, I_3
 \end{equation}
 with $I_3$ varying from 0 to 1. The full $I_3$ expression is given in equations 1 and 4 in \cite{bie15} and  in Eqs.\,5--8,
\begin{equation}
\tilde{\sigma}_R=  \tilde{\sigma}_{0,R} \exp \left( -R_c(L_z)/ R_{\sigma_R} \right) \, ,
\end{equation}
\begin{equation}
\tilde{\sigma}_z=  \tilde{\sigma}_{0,z} \exp \left( -R_c(L_z)/ R_{\sigma_z} \right) \, ,
\end{equation}
 \begin{equation}
g(L_z)=
\frac{ 2 \, \Omega(R_c)  }
{\kappa(R_c) } \, 
\frac{ \tilde{\rho}(R_c)  \, }
{(2 \pi)^{3/2} \,    \tilde{\sigma}_R(R_c)^2 \, \tilde{\sigma}_z(R_c)} \, ,
 \end{equation} 
\begin{equation}
\tilde{\rho}=  \tilde{\rho}_{0} \exp \left( -R_c(L_z)/ R_{\rho} \right) \, .
\end{equation}

The parameters $\tilde{\sigma}_R$ and  $\tilde{\sigma}_z$ allow us to define the radial and vertical velocity dispersions that have   nearly exponential radial decreases  with the scale lengths  $R_{\sigma_R}$ and $R_{\sigma_z}$. The parameter $\tilde{\rho}$ defines the density distribution that is also  nearly exponential with the scale length $R_{\rho}$. The parameters $\tilde{\rho}_{0}$,   $\tilde{\sigma}_{0,R}$, and   $\tilde{\sigma}_{0,z}$  scale  the global amplitudes of the density and velocity dispersions. The thicknesses of these discs are no longer free parameters and they are constrained by the $\tilde{\sigma}_z$ and the $R_{\sigma_R}$ parameters.  We write each of  the free model parameters with a tilde. For small velocity dispersions, the exact velocity dispersions $\sigma_{R}$ or $\sigma_{z}$ are close to the functions  $\tilde{\sigma}_{R}$  or $\tilde{\sigma}_{z}$, and the exact density distribution ${\rho}$ is close to $\tilde{\rho}$. The parameters $\Omega$ and $\kappa$ are the circular velocity  and the epicyclic frequency.
\\

If $I_3=0$ or $E_z=0$, the DF (Eq. 9) reduces to the Shu DF with  the density  null outside of the mid-plane $z=0$. At the opposite, and  for St\"ackel potentials, if $I_3$=1 (i.e. $E_R$=0), the corresponding orbits are the shell orbits. The volume occupied by shell orbits in a St\"ackel potential is two-dimensional and these orbits are confined in an ellipsoidal sheet with no radial extension when they cross the plane at $z=0$.

In the case of non-St\"ackel potentials, as the BGM potential,  the  approximate third integral $I_3$  of the shell-like orbits reaches a maximum value, $I_{3,max}(E,L_z)$,  that is different from 1   \cite[see e.g.][figure 31 for  extreme case of a thick shell orbit]{oll62}. Then, in the case of non-St\"ackel potentials  the DFs for stellar discs must be written using equation Eqs. 9--11 and replacing Eq. 10 with
\begin{equation}
E_R= (E-E_c) \, (I_{3,max}-I_3) \, ,
\end{equation}
$E_R$ specifies the amount of radial motion and $E_z$ the amount of vertical motion.

Finally, we  note that the DF (Eq. 9) is quasi isothermal, and exactly  isothermal  in the case of a separable potential in $R$ and $z$ coordinates. It is  similar to the DFs used by \citet{bin11}  since in the  case of  small departures from circular motions we have
\begin{equation}
 E_R \simeq \kappa J_R
\end{equation}
and 
\begin{equation}
 E_z \simeq \nu J_z \, ,
\end{equation}
where $\nu$ is the vertical frequency, $J_R$ and $J_z$ the radial and vertical actions.

Finally,  we introdruce a cut in the DF (Eq. 9), with $g(L_z) = 0$ if  the  angular momentum is larger than $L_{z,cut}$=15 kpc $\times$ 200 km/s. This cut  models the decrease in the stellar disc density distributions at large Galactic radii \citep{amo17}

\section{Dynamically consistent model}

The previous paragraphs detail the elements needed to build a dynamically consistent Galactic model and  each stellar disc  is characterised with a DF given by Eq. 9. These DFs are computed for an initially given potential.  Since they generate new densities and a new potential, we  iterate until obtaining the convergence of the potential and of the DFs. The resulting Galactic model depends on the list of the following free parameters $\tilde{\rho}_0$, $\tilde{\sigma}_{0,R}$, $\tilde{\sigma}_{0,z}$, $\tilde{R}_{\rho}$, $\tilde{R}_{\sigma_R}$, and  $\tilde{R}_{\sigma_z}$ given for each stellar disc (i.e. the densities and velocity dispersions at the solar position and their respective scale lengths).

We apply the process to  adjust the  free parameters and fit  the analytical density of each stellar disc of the BGM2014 version (Einasto profiles and modified double-exponential profiles). The fit is  restricted radially to the domain  4 kpc $< R <$ 12 kpc  to avoid the modelling of the central hole of the BGM2014 analytic discs.
The fit is also restricted vertically to a maximum z-height at  3 to 4 scale heights for each  considered disc. Moreover for the thick disc, we restrict the fit to $z$-distances smaller than 1 kpc. These $z$-limits are introduced since  within the 2004 BGM version there was no attempt to perform the dynamical consistency at larger $z$-distances. Our adjustment is achieved by a least-squares minimisation of the density profiles in the given ranges of $R$ and $z$ positions  by  using the MINUIT software \citep{jam04}

The quality of the fit of the density distributions (Figures 1 and 2)  is  excellent,  revealing  that the family of stellar density profiles were  realistically chosen in \citet{rob86a,rob86b}.  
 However in these models, the kinematics, which are    dynamically consistent only in the solar neighbourhood,  had to be improved  and this is  achieved with the dynamical modelling conducted here. This new modelling suppresses  free parameters such as the velocity dispersion ratios $\sigma_R/\sigma_\theta$ and  the disc scale heights. Furthermore,  the asymmetric drift is determined exactly and is not   approximated.

With the new version of the model,    adjusting the kinematical parameters $\tilde{\sigma}_{z}$ and  $\tilde{R}_{\sigma_z}$  and  the density parameters $\tilde{\rho}_{0}$ and  $\tilde{R}_{\rho}$ is the predominant factor that allows us to     reproduce the BGM2014 densities of the stellar discs. Conversely, varying $\tilde{\sigma}_R$ and $\tilde{R}_{\sigma_R}$  within reasonable values has  a small impact on  the vertical and radial density distributions of the stellar discs. These two last parameters will be  constrained with observations and not by the dynamical self-consistency. Thus, these two parameters $\tilde{\sigma}_R$ and $\tilde{R}_{\sigma_R}$  are currently  fixed before they can be more tightly constrained by kinematical observations. Then we assume that $\tilde{R}_{\sigma_R}= \tilde{R}_{\sigma_z}$ and we fixed for each stellar disc
 the velocity dispersion ratio  ${\sigma}_U/{\sigma}_W$ at the solar position, respectively 2.1 for the thin discs \citep{gom97,rob17}, 1.47 for the young thick disc, and 1.38 for the old thick disc \citep{sou03,rob17}.
\\

\begin{figure*}[!htbp]
\begin{center}
\resizebox{7.1cm}{!}{\rotatebox{270}{\includegraphics{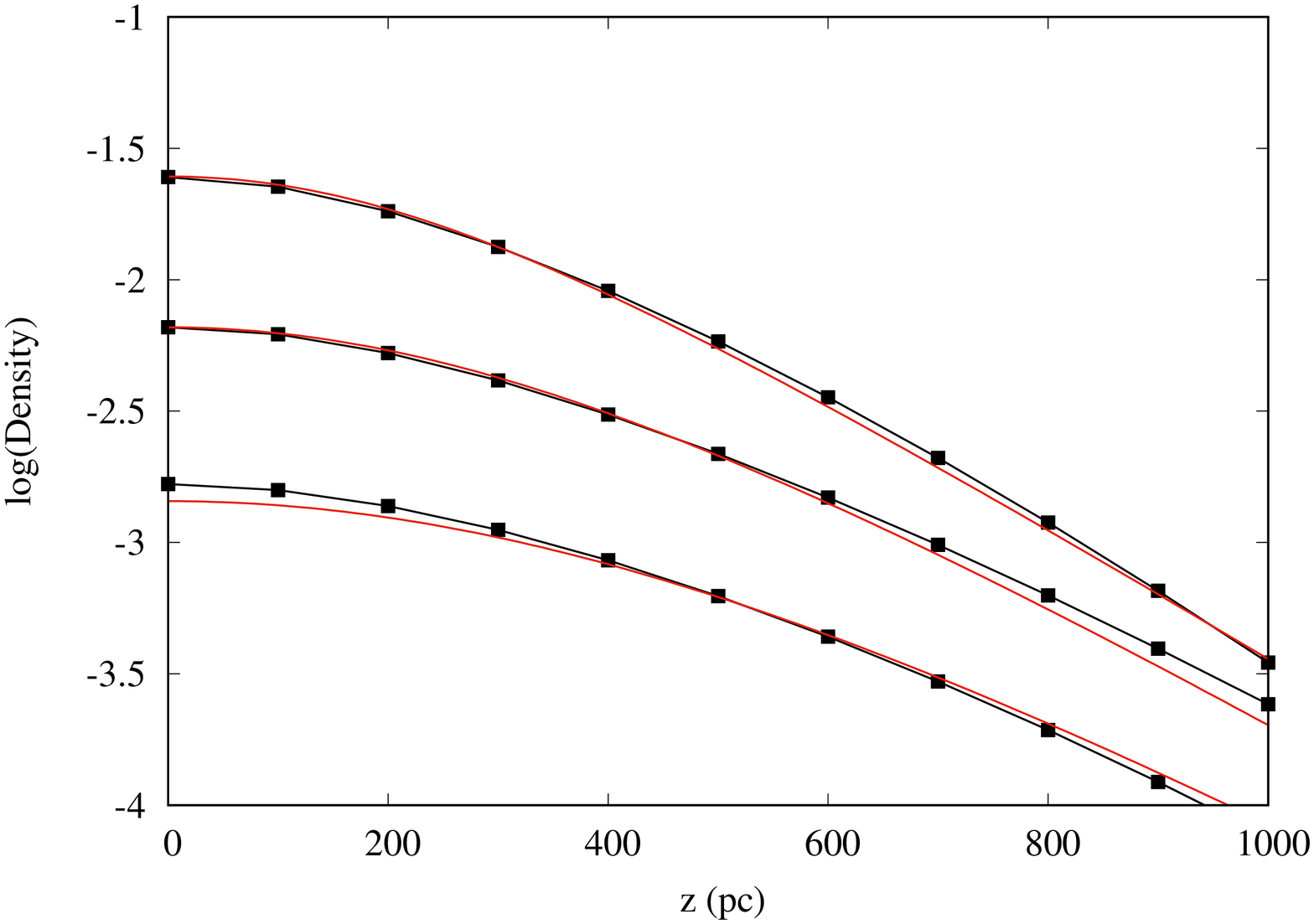}}}
\resizebox{7.1cm}{!}{\rotatebox{270}{\includegraphics{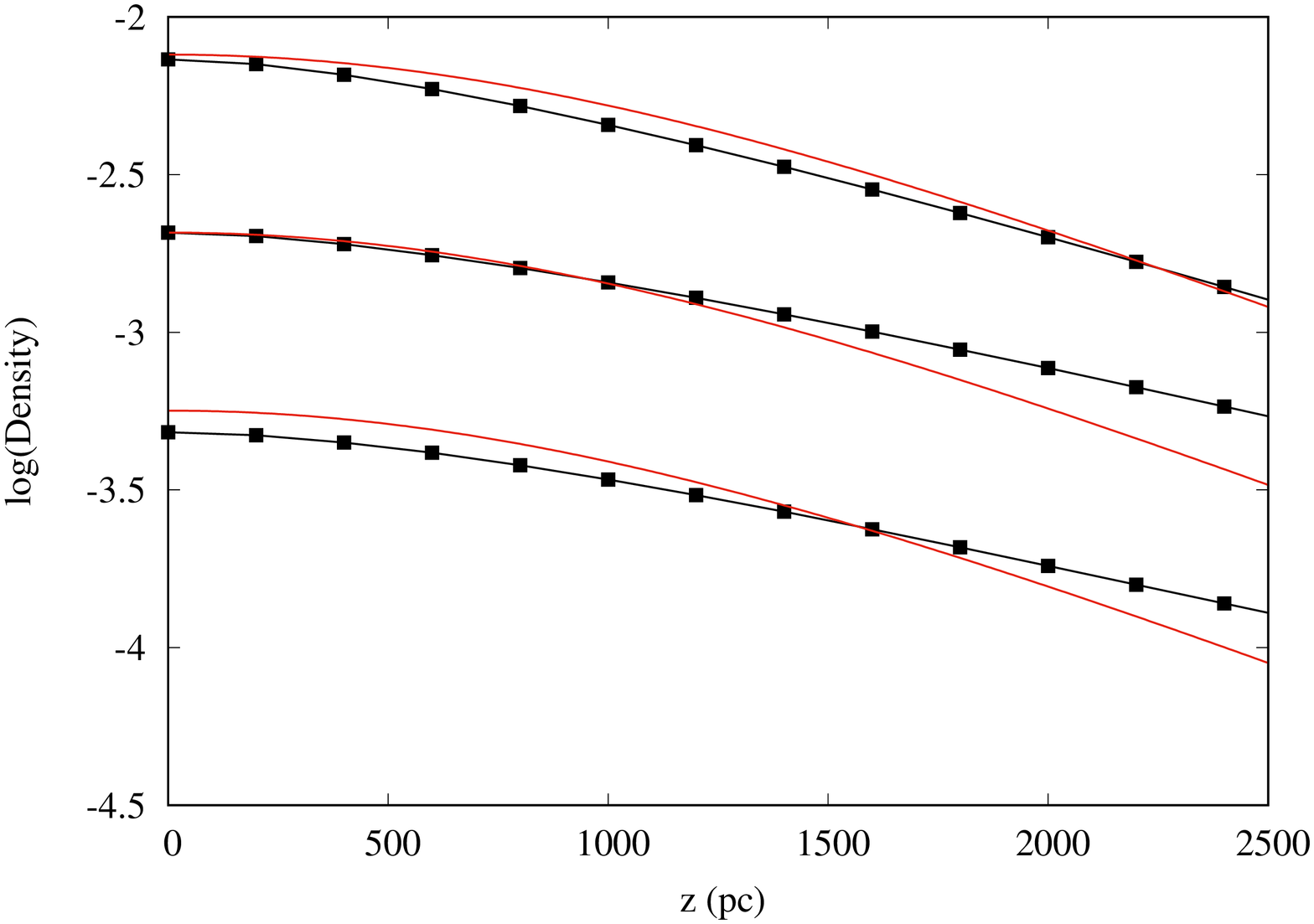}}}
\end{center}
  \caption{ Vertical density distributions  for the old thin stellar disc (left) and the old thick disc (right)  at three Galactic radii $R$ = 4, 8, and 12 kpc.  Red lines: the old thin disc Einasto profile (left) and the old thick disc profile (right)  from   the BGM2014 version. Black squares and lines:  dynamically consistent fitted discs.
  }
    \label{fig1}
\end{figure*}

\begin{figure*}[!htbp]
\begin{center}
\resizebox{7.1cm}{!}{\rotatebox{270}{\includegraphics{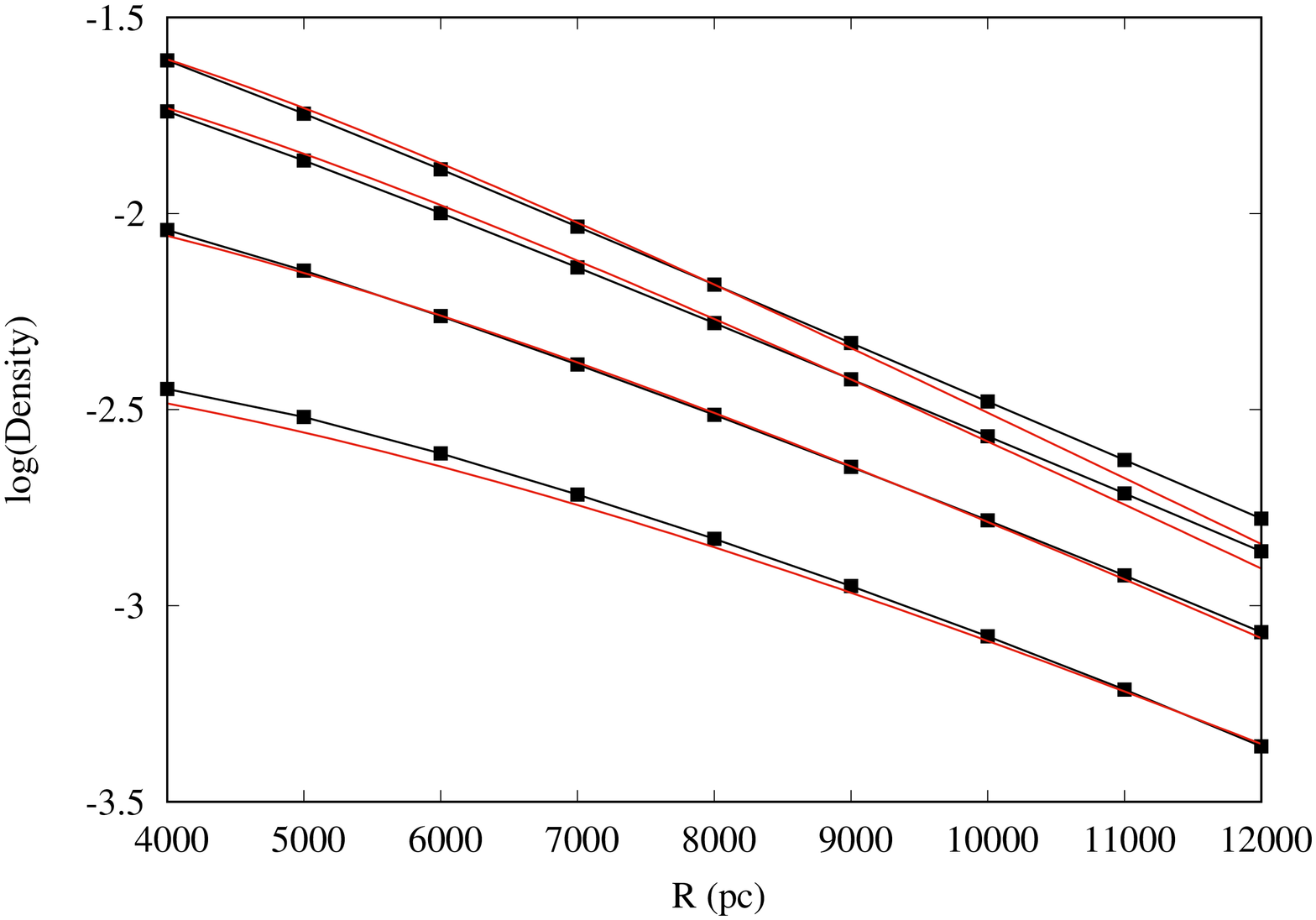}}}
\resizebox{7.1cm}{!}{\rotatebox{270}{\includegraphics{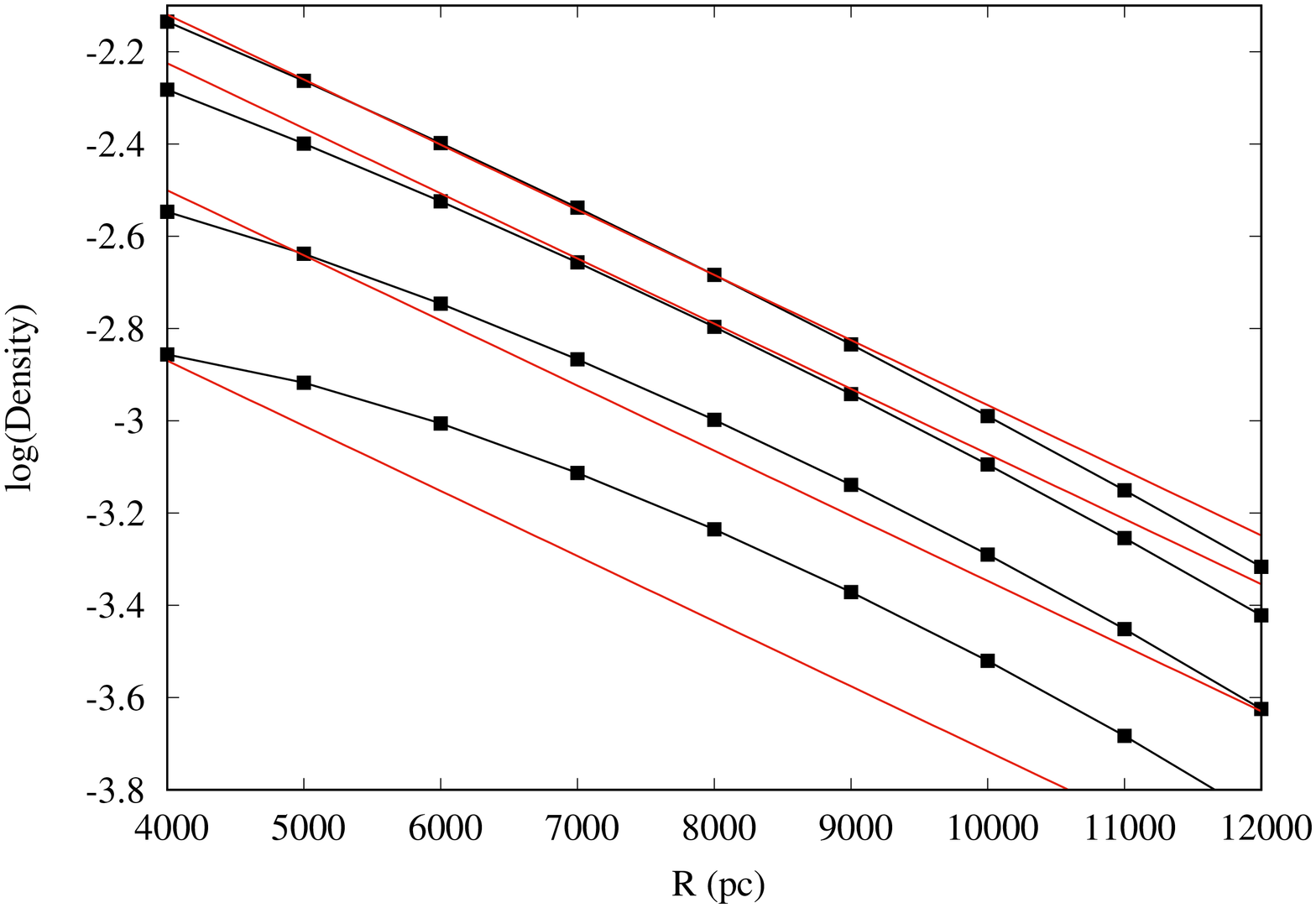}}}
\end{center}
  \caption{ Radial density distributions  for  the old thin stellar disc (left) and the old thick disc (right)  at four vertical heights $z$.  Red lines: the old thin disc Einasto profile at $z$ = 0, 200, 400, and 600 pc (left) and the old thick disc profile  at $z$ = 0, 0.8,  1.6,  and 2.4  kpc (right) from the BGM2014 version.  Black squares and lines:  dynamically consistent disc  fitted from $R$ = 4 kpc to 12 kpc. 
  }
    \label{fig2}
\end{figure*}

Figure 1  (left) shows the vertical density distribution of the old thin disc (scale height around 230 pc) at three Galactic radii 4, 8, and 12 kpc, and Figure 2 (left) its radial distribution at four $z=$ 0, 200, 400, and 600 pc. The adjustment is satisfying. This is expected at $R$ = 8 kpc since the BGM2014 version is built to be dynamically consistent up to $z$ = 1 kpc. The fact that the fit is also satisfying at  other galactic radii  confirms  that the choice of  Einasto profiles  in the BGM2014 version  for the disc density was valuable with respect to the dynamics. 
 For the other  discs,  the agreement between the BGM2014 and the new version is correct and is better for the thinner disc components with smaller vertical velocity dispersions. For the old thick disc, the vertical and radial density distributions  (corresponding to a  795 pc scale height  in the case  of a $sech^2$ density law) are shown in Figures 1 and 2 (right).  The agreement between both models is still correct. We note that at $R = 8$ kpc and below $z$ = 1 kpc the fit is excellent, illustrating    the exactness of the dynamical consistency of the BGM2014 version. However, at larger height above the Galactic plane, both models differ, and at two scale-heights, or 1.6 kpc height, the BGM2014 density for the old thick disc is 17 \%\  too small.

Concerning the density scale length, the agreement between the two models is nearly perfect for the old thin disc. For the old thick disc, the agreement is obtained only for $z$ lower than 1 kpc. We note that the flare present in the outskirts of the Galaxy could be naturally modelled by adjusting the vertical velocity dispersions at large Galactic radii.

Table 1 gives the   relative difference  in percentage between the new and the old models at different heights $z$ and at $R$ = 8 kpc, for the old thin and old thick discs. The differences increase with $z$; they become significant at three scale heights for the old thin disc and beyond 1 kpc for the old thick disc. At  high $z$, the differences become of the order of the expected density of the stellar halo, and  this illustrates the necessity of using dynamically self consistent distribution functions to properly identify and separate the Galactic stellar components.
However, only observations will tell us whether our choice of nearly isothermal modelling of stellar discs is the correct one since   small changes in the wings of the velocity distribution at low $z$ impact the vertical density distribution at high $z$. Other choices of DFs could produce  similar densities and velocities  at low $z$ and different densities at high $z$.

\begin{table}[htp]
\caption{
Density differences (in percentage) between the BGM2004 version and the new dynamically self-consistent model for the old thin and old thick discs at $R$ = 8 kpc and various $z$.
}
\begin{center}
\begin{tabular}{|c|c|c|c|c|c|c|c|c}
\hline
old thin disc &&&& \\
\hline
z (pc) & 200 & 400 & 600 & 800 \\
\hline
 diff(\%) &  -2.3\%& -1.1\%& +5\%& +13\% \\
 \hline
 old thick disc &&&&\\
\hline
z (pc) &  800 & 1600 & 2400 & 3200 \\
\hline
 diff(\%) &  -2.5\%& +17\%& +60\%& +263\% \\
 \hline
\end{tabular}
\end{center}
\label{default}
\end{table}%

The vertical velocity dispersion of each stellar component at ($R_{\,0}$, $z$ = 0) is  marginally modified in the new version compared with the BGM2014 version (see Table 2). 
This is a consequence of the fact that  the self-consistency was already applied at low\,$z$.
\begin{table}[htp]
\caption{Previous \citep{rob14,rob17} and new vertical velocity dispersions in km.s$^{-1}$ for disc 2 to 7  components and the two young and old thick disc components (8 and 9)
at $R_0$= 8 kpc and $z$= 0.}
\begin{center}
\begin{tabular}{|c|c|c|c|c|c|c|c|c|}
\hline
disc & 2 & 3 & 4 & 5 & 6 & 7 & 8 & 9\\
\hline
previous & 8 & 10 & 13.2 & 15.8 & 17.4 & 17.5  & 28 & 59 \\
%{\it Holmberg} & {\it 8} & {\it 10} & {\it 14} & {\it 17.5} & {\it 21.0} & {\it 25.0} \\
new   &  7.3 & 9.9 & 13.4 & 16.0 & 17.5 & 17.6 & 26.4 & 51 \\
\hline
\end{tabular}
\end{center}
\label{default}
\end{table}%

\begin{figure}[!htbp]
\begin{center}
\resizebox{7.1cm}{!}{\rotatebox{270}{\includegraphics{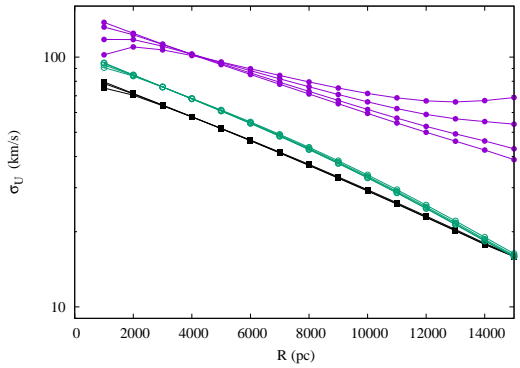}}}
\end{center}
  \caption{ $\sigma_U$ radial velocity dispersions for the old thin disc (black lines)  
  and  the young thick disc (green lines) at $z$ = 0, 200, 400, and 600 pc. For the old thick disc (purple lines)  at $z$ = 0, 0.8,  1.6,  and 2.4  kpc.  We note an increase  in $\sigma_U$ and $R_{\sigma_U}$  with $z$  for the old thick disc.
    }
    \label{fig3}
\end{figure}

\begin{figure}[!htbp]
\begin{center}
\resizebox{7.1cm}{!}{\rotatebox{270}{\includegraphics{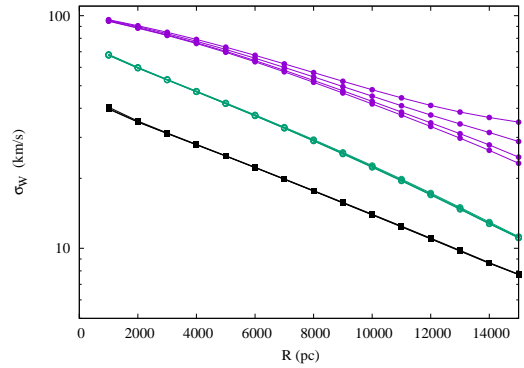}}}
\end{center}
  \caption{ $\sigma_W$ vertical  velocity dispersions for the old  thin disc  (black lines)  and the young thick disc (green lines) at $z$ = 0, 200,  400, and 600 pc. For the old thick disc (purple lines)  at $z$ = 0, 0.8,  1.6,  and 2.4  kpc, $\sigma_W$ does not change with $z$ for the old thin disc or the young thick disc, but  for the old thick disc,  $\sigma_W$, and $R_{\sigma_W}$  increase with $z$. 
    }
    \label{fig4}
\end{figure}

Figure 3 shows at different $z$  the radial  variations  of  $\sigma_U$ and Figure 4  the  variations of $\sigma_W$. The velocity dispersions $\sigma_U$ and $\sigma_W$ are the major and minor axes of the velocity ellipsoid (i.e. $\sigma_R$ and $\sigma_z$ if $z=0$).  By construction, their radial  variations are nearly exponential.  For the old thin disc, we do not see  significant $z$-variation of the  kinematical scale lengths,  $R_{\sigma{_R}}$ and $R_{\sigma{_z}}$. For the old thick disc, the kinematic scale lengths increase with $z$.  For the thin and thick discs discussed here, the respective density scale lengths  $R_\rho$ at $z=0$ are 2.53 and 3.15 kpc, and the kinematic scale lengths $R_{\sigma_z}$ are 8.5 and 10.8 kpc. The ratio $R_{\sigma_z}/R_\rho$ is far from the factor 2,  a frequently assumed value  based on the hypothesis of a self-gravitating isothermal disc  \citep{van81,van82}.
At very small radii $R$, the asymmetric drift is very large for the old thick disc  and moreover the DFs are null for negative angular momentum. Thus, the DFs are  not realistic and it could explain the decrease in the dispersion at low $R$ and high $z$   (see Fig. 3).

The asymmetric drifts and the $\sigma_V$ velocity dispersions are no longer free parameters and are exactly determined.  The asymmetric drifts (Figure 5) are shown as a function of $z$ at the solar Galactic radius $R_0$. The very different asymmetric drift of the old thick disc is due to its much higher velocity dispersions. 

Within the interval $R=$ 4 to 12 kpc,  the velocity ellipsoids roughly point  towards the Galactic centre.  Figure 6 shows the variation in the tilt with $z$ at $R$ = 8 kpc for all the stellar components.    They are very similar and just the old thick disc has a  tilt slightly larger  by  1 degree at 4 kpc height. 
  The velocity ellipsoid tilt variation for  the stellar discs  results from the  different $z_0$ values defining the St\"ackel coordinates to model the DFs. These different $z_0$ values are directly obtained by fitting the orbits populating the stellar discs.
  We can consider as a first approximation that the tilt does not depend on the stellar disc population.

\begin{figure}[!htbp]
\begin{center}
\resizebox{7.1cm}{!}{\rotatebox{270}{\includegraphics{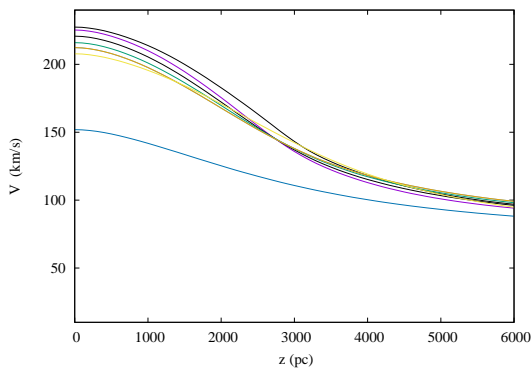}}}
\end{center}
  \caption{ Mean circular velocities  for all  disc components vs. $z$ at $R$ = 8 kpc with $V_c(R_\sun)$ = 221 km/s. The blue line corresponds to the old thick disc.
    }
    \label{fig5}
\end{figure}

\begin{figure}[!htbp]
\begin{center}
\resizebox{7.1cm}{!}{\rotatebox{270}{\includegraphics{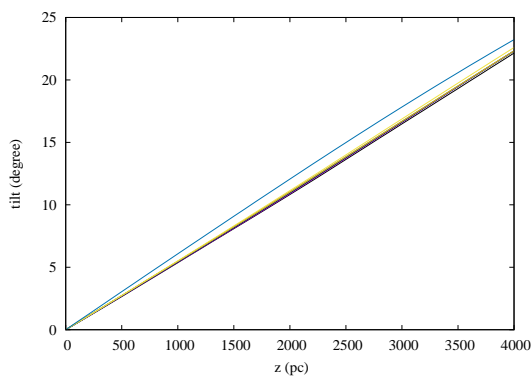}}}
\end{center}
  \caption{ Velocity ellipsoid vertical tilts  for stellar discs   vs. $z$ at $R$ = 8 kpc. The blue line corresponds to the old thick disc.
    }
    \label{fig6}
\end{figure}

\section{Conclusion}

We have presented the elements  of the structure of the new  dynamical self-consistency of  the Besan\c{c}on Galactic Model. It is  based on the use of an explicit approximate third integral and on the 3D generalisation of the Shu DF. For each stellar disc, the number of free parameters is reduced since  the vertical velocity dispersions and the scale heights are dynamically related for each stellar disc. In addition  the density scale heights and  the vertical kinematic scale lengths ($R_{\sigma_z})$ are linked. A preliminary and intermediate version of this model was used to adjust the stellar kinematics from RAVE and TGAS observations \citep{rob17}.\\

The consistency of the dynamical model is performed with an accuracy that allows us to recover the mass distribution and forces with  an error smaller  than a few thousandths. Since, this new model substantially reproduces the previous 2004 version of the Besan\c{c}on model and its stellar densities, in consequence it also matches the star counts and  then     provides a prediction of the kinematics of stellar populations.   These predictions will be compared with Gaia observations.   We can expect that these new observations will   entail revisions with an adjustment of the modelled stellar populations, and an improvement of our current representation of the dark matter distribution. 

A precise determination of the mass distribution of the Galaxy and the dark matter component will only be only achieved  with an accurate modelling of the observed stellar components and their kinematics. This is  an immediate goal of the BGM. Furthermore, obtaining a stationary  model   of the stellar discs will  be a preliminary step to estimate the degree of non-stationarity of  stellar components. For each stellar population, the   accuracy of the fit will give  the amplitude of non-stationarities. This will help to identify what type of non-stationarities are involved and  what kind of complementary models must be developed.

To conclude, we note that we have considered a stationary Galaxy model without a bar and our methods require axisymmetric potentials. Generalisation to a non-axisymmetric model (work in preparation) is however possible with the same simplicity since the method developed for axisymmetric potentials \citep{bie15} can be  generalised to 3D, also with     analytic approximate integrals explicitly depending on the potential without a need for numerical integrations
\cite[see also][]{saneva15,sanbin15}.

\begin{acknowledgements}
We are grateful to the anonymous referee for his report which improved the manuscript.
\end{acknowledgements}

%\bibliographystyle{aa} % style aa.bst
%\bibliography{} % YOUR REFERENCES WD.bib

% for the bibliography, at the end  
\bibliographystyle{aa} % style aa.bst 
\bibliography{MDB18} % your references Yourfile.bib
%\bibliography{VersionNov3} % your references Yourfile.bib

\end{document}